\documentclass[twocolumn,showpacs,preprintnumbers,amsmath,amssymb]{revtex4}
\usepackage{graphicx}
\usepackage{dcolumn}
\usepackage{bm}
\begin{document}
\title{Noise, Synchrony and Correlations at the Edge of Chaos}

\author{Alessandro Pluchino $^{1,2}$, Andrea Rapisarda$^{1,2}$ and Constantino Tsallis $^{2,3}$}
\bigskip 

\affiliation{$^1$ Dipartimento di Fisica e Astronomia, 
Universit\`a di Catania  and INFN sezione di Catania, 
Via S. Sofia 64, 95123 Catania, Italy\\
$^2$ Centro Brasileiro de Pesquisas Fisicas and National Institute of Science and Technology for Complex Systems
22290-180 Rio de Janeiro-RJ, Brazil \\
$^3$ Santa Fe Institute - Santa Fe, NM 87501, USA	}
\date{\today}

\begin{abstract}

We study the effect of a weak random additive noise in a linear chain of N locally-coupled logistic maps at the edge of chaos. Maps tend to synchronize for a strong enough coupling, but if a weak noise is added,  very  intermittent fluctuations in the returns time series are observed. This intermittency tends to  disappear when noise is increased. Considering the pdfs of the returns, we observe the emergence of fat tails which can be satisfactorily reproduced by $q$-Gaussians curves typical of nonextensive statistical mechanics. Interoccurrence times of these extreme events are also studied in detail. Similarities with recent analysis of financial data are also discussed. 

\end{abstract}
\maketitle
\vspace{0.25cm}

Since the nonlinear phenomenon of synchronization was first observed and discussed in the 17th century by Huygens, it has become of  fundamental importance in various fields of science and engineering. It is frequently  observed in complex systems such as biological ones, or  single cells, physiological systems, organisms and even populations.  Synchrony among coupled units has been extensively studied in the past decades providing important insights on the mechanisms that generate such  an emergent collective behavior \cite{Strogatz,Pikovsky,Bonilla,Kuramoto1,Miritello}. In this context coupled maps have  often been used  as a theoretical  model \cite{Kaneko1}.  Actually, many biological complex systems operate in a noisy environment and most likely at the edge of chaos \cite{Langton,Kauffman}. Therefore studying the effect of a weak   
noise in this kind of coupled systems could be relevant in order  to understand the way in which interacting units behave in real complex systems like for example living cells \cite{Stefan1,Stefan2}. 
Generally speaking, random noise is considered a disturbance, i.e. something to avoid in order to obtain precise measurements or to minimize numerical errors. But, if on one hand it is very difficult to completely eliminate the effect of noise, on the other hand it can frequently have even a beneficial role. Among the many examples in physics and biology we may cite stochastic resonance \cite{Gammaitoni}, noise enhanced stability \cite{Mantegna}, induced second-order like phase transitions \cite{Parrondo1} or  enhanced diffusion in communication networks  \cite{Caruso}. Recently random strategies have been demonstrated to be very successful also in minority and Parrondo games \cite{Parrondo2,Sornette1} and in sociophysics models related to efficiency in hierarchical organizations  \cite{Pluchino1,Pluchino2} or  even in Parliament models \cite{Pluchino3}. 
\begin{center}
\begin{figure}
\includegraphics[width=1.92in,angle=0]{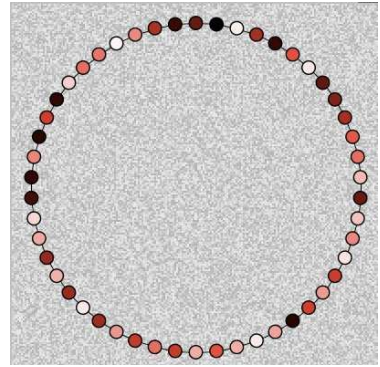}
\caption{\label{fig1} 
(Color online) A pictorial view of a chain of $N=50$ locally coupled logistic maps embedded in a noisy environment. The different colors indicate, at a fixed  time $t$,  different values of the maps in the interval $[-1,1].$  
}
\end{figure}
\end{center}
In previous  studies \cite{Kaneko3,Kuramoto2} the effect of a small noise on globally coupled chaotic units was presented for several kind of systems and a universal behavior related to the Lyapunov spectrum was found to be a common feature. Power-law correlations and intermittent behavior have  also been observed in lattices of logistic maps when some kind of global coupling exist among them, see for example refs.\cite{Kaneko2,Li-Fang}. 
In this letter,  we consider  a linear  chain of N locally coupled logistic maps  and we explore  the role that a small random noise can have in creating  a strong intermittent behavior and its influence on the  synchronization patterns.
We consider only local coupling and long-range correlations induced by the noisy environment in which  our maps are embedded. At variance with previous studies, maps  are not in a  chaotic regime, but at the edge of chaos, where the Lyapunov exponent is vanishing \cite{Robledo,Beck2}. Moreover, in order to investigate our intermittent behavior, we study the pdfs of the  returns of  our fluctuating time series, as successfully done in  different contexts for models showing Self-Organized Criticality  \cite{Caruso2,Ugur1,Ugur2}.  In this respect our model presents new features never addressed before as far as we know and more related to realistic complex systems.
Our results can also be  framed in the context of non extensive statistical mechanics \cite{Tsallis,Hmf1,Hmf2} and  analogies with recent findings   for stock market data analysis  \cite{Bunde,Bunde2} will be addressed. 
%
\begin{center}
\begin{figure}
\includegraphics[width=3.2in,angle=0]{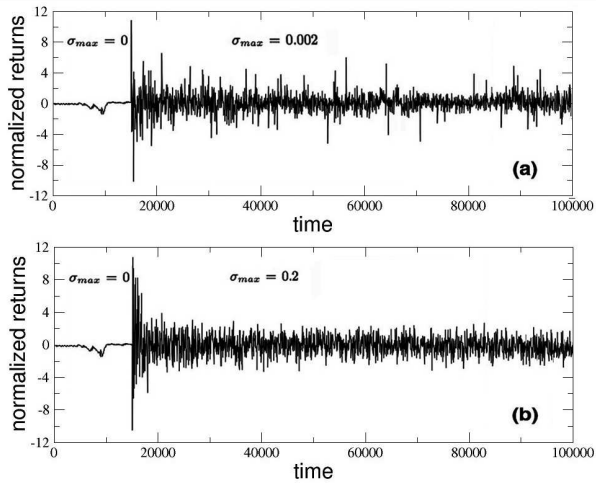}
\caption{\label{fig2} 
We show the effect of noise in the normalized returns of Eq. (2) for the case $N=100$, $\mu=\mu_c=1.4011551...$, $\epsilon=0.8$  and $\tau=32$ time steps. 
At time $t=15000$ we switch on the noise, with $\sigma_{max}=0.002$ in panel (a) and $\sigma_{max}=0.2$ in panel (b), then we follow the maps for $100000$ iterations.  
See text for further details.
}
\end{figure}
\end{center}
The model  of a linear chain of N   coupled logistic maps  is the following 
\begin{equation}
x_{t+1}^i = \left(1- {\epsilon } \right) f \left({x_t}^i \right) + { \frac{\epsilon} {2} }  \left[ f \left( {x_t}^{i-1} \right)
+ f \left({x_t}^{i+1} \right) \right]   + {\sigma(t)}
\label{coup}
\end{equation}
where $ \epsilon  \in \left[ 0,1 \right] $
is the strength of the local coupling of each map with its first neighbors sites on the chain and the additive noise $\sigma(t)$ is a random variable, fluctuating in time but equal for all the maps, uniformly extracted in the range $\left[ 0, \sigma_{max} \right]$. In our case  the $i$-th logistic map at time $t$ is in the form  $ f \left(x_t^i \right) = 1 - \mu \left( x_t^i \right)^2 $, 
with $\mu \in [ 0, 2 ]$ and with $f \left(x_t^i \right)$ taken in module $1$ with sign (in order to fold the maps' outputs back into the $[-1,1]$ interval when the noise takes them out of it). The system has periodic boundary conditions. See Fig.1 for a pictorial view. 
\\
In the absence of noise, this model was extensively studied by Kaneko et al. \cite{Kaneko1}, in particular in the chaotic regime, where the coupled maps show different patterns of synchronization as function of the coupling strength $\epsilon$. Here we consider the effect of a variable addition of noise on the same system, but at the edge of chaos, i.e. at  $\mu_c=1.4011551...$ . Following a procedure adopted in ref.\cite{Li-Fang}, in order to subtract the synchronized component and keep the desynchronized part of each map we consider, at every time step, the difference between the average $< x_t^i >$ and the single map value $x_t^i$.
\begin{center}
\begin{figure}
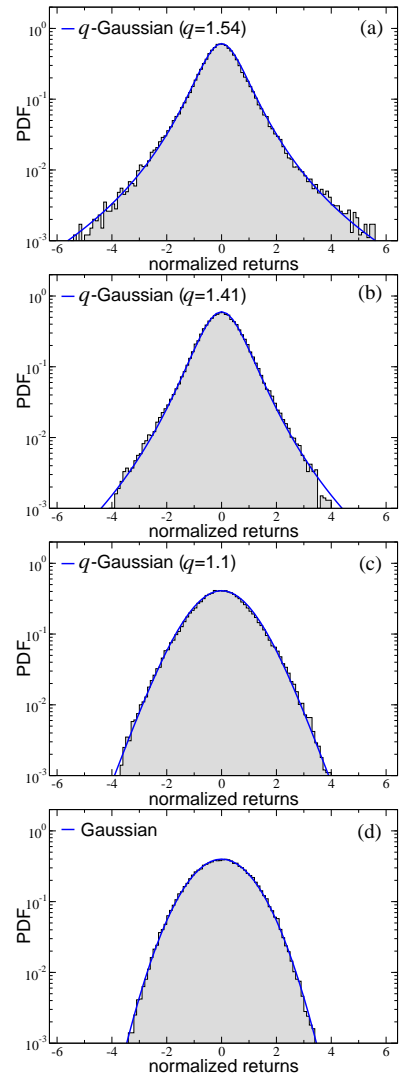

\includegraphics[width=2in,angle=0]{Fig3a.eps}
\includegraphics[width=2in,angle=0]{Fig3b.eps}
\includegraphics[width=2in,angle=0]{Fig3c.eps}
\includegraphics[width=2in,angle=0]{Fig3d.eps}
\caption{\label{fig3} 
Asymptotic pdfs of the returns for $N=100$ maps at the edge of chaos, i.e. $\mu_c=1.4011551...$, with $\epsilon=0.8$, $\tau=32$ (after 100000 iterations) and for increasing values of the noise: $\sigma_{max}=0.002$ (a), $\sigma_{max}=0.01$ (b), $\sigma_{max}=0.05$ (c) and $\sigma_{max}=0.3$ (d). Fat tails are more evident for weak noise and tend to diminish by increasing noise. We report also fits of the simulation data (full curve) by means of q-Gaussian curves with values $q=1.54$, $q=1.41$, $q=1.1$ and $q=1$ (corresponding to a Gaussian) respectively. Returns are also normalized to the standard deviation in order to have pdf  with unit variance.  See text for further details.  
}
\end{figure}
\end{center}
Then we further consider the average of the absolute values of these differences over the whole system, i.e. $
d_{t} =  \frac{1} {N} \Sigma_{i=1}^N  | x_t^i  -  < x_t^i > |
$,
in order to measure the distance from the synchronization regime at  time $t$. If all maps are trapped in some synchronized pattern then this quantity remains close to zero, otherwise oscillations are found. As commonly used in turbulence or in finance \cite{Tsallis,Bunde,Bunde2}, we analyze these oscillations by considering the two-time returns $\Delta d_{t}$ with an interval of $\tau$ time steps, defined as  
$
\Delta d_{t} =  d_{t + \tau}  - d_{t } 
$.
\\
In Fig.2 we show that this quantity is very sensitive to the noise intensity.
In panels (a) and (b) we plot the time evolution of $\Delta d_{t}$ (normalized to the standard deviation of the overall sequence) for two different simulations obtained with a linear chain of $N=100$ maps, with $\epsilon=0.8$, $\tau=32$ and considering the maps at the edge of chaos. For both the simulations we consider a transient of $15.000$ iterations, during which the system evolves in the absence of noise ($\sigma_{max}=0$), then we suddenly increase the level of noise bringing it on at $\sigma_{max}=0.002$ (a) and $\sigma_{max}=0.2$ (b) respectively: it clearly appears that only in presence of weak noise (panel (a)) the returns time series shows large deviations from the synchronized pattern of the transient, while a higher noise intensity destroys the intermittency and induces Gaussian fluctuations.
\\
%
\begin{center}
\begin{figure}
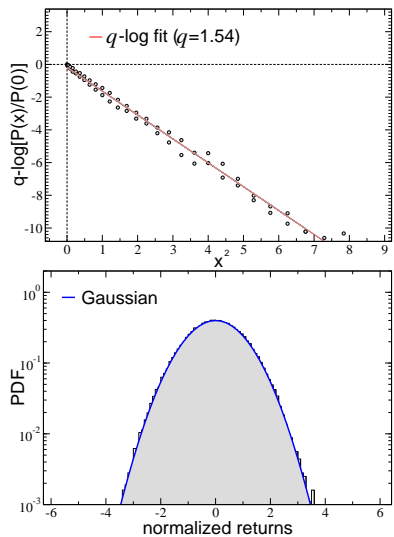

\includegraphics[width=2in,angle=0]{Fig4a.eps}
\includegraphics[width=2in,angle=0]{Fig4b.eps}
\caption{\label{fig4} 
Top panel: The $q$-logarithm of the pdf reported in Fig.3(a) (normalized to the peak) is plotted as function of $x^2$. A q-logarithmic curve with $q=1.54$ fits the points  with  a correlation coefficient equal to 0.9958.
Bottom panel: Same simulation reported  in Fig.3(a), but  with the maps in the fully chaotic regime ($\mu = 2$).
See text for further details.  
}
\end{figure}
\end{center}
%
\begin{center}
\begin{figure}
\includegraphics[width=3.4in,angle=0]{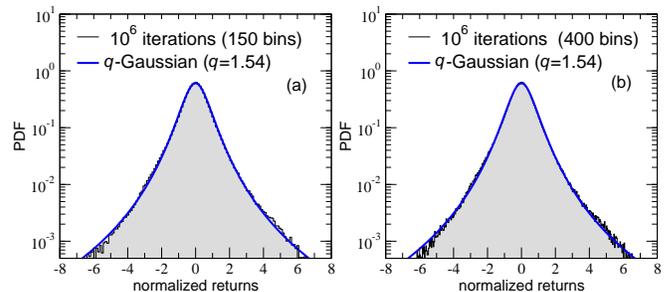}
\caption{\label{fig5} 
Asymptotic pdfs of the returns for $N=100$ maps at the edge of chaos, i.e. $\mu_c=1.4011551...$, with $\epsilon=0.8$, $\tau=32$  as in the case of Fig.3(a), but now taken with higher statistics in order to test the robustness of previous results. In this case  we considered   $10^6$ iterations and two histograms with a different number of bins, i.e.  (a) 150 and (b) 400 respectively.  The same $q$-Gaussian curve reported in Fig.3(a),   here shown  for comparison, reproduces very well the new numerical simulations independently of the size of the bins used for the histogram.  
}
\end{figure}
\end{center}
In order to quantify such a different noise-dependent behavior we plot in Fig.3 the asymptotic probability distribution function (pdf) of normalized returns for increasing values of $\sigma_{max}$, from  $0.002$ to $0.3$.  Fat tails in the pdfs are visible only for $\sigma_{max}<0.1$ and, following what was already done for a single logistic map at the edge of chaos \cite{Robledo,Beck2}, we tried to reproduce them by q-Gaussian curves, typical of non extensive statistical mechanics \cite{Tsallis} and usually found in complex systems presenting various kinds of correlations. q-Gaussians are defined as 
$
G_q(x)= A\left[1- \left(1-q \right) \beta x^2 \right]^{\frac{1}{1-q}},
$
where $q$ is the entropic index (which evaluates deviations from Gaussian behavior), $1/\beta$ plays the role of a variance and $A$ is a normalization parameter. These curves actually fit very well the numerical pdfs, as also reported in the panels of Fig.3 (full lines). For $\sigma_{max}=0.002$ (panel (a)), where the tails are very pronounced, one has $q\sim1.5$ while, for higher values of noise, the tails tend to disappear and the value of $q$ decreases asymptotically towards $q=1$, which corresponds to a Gaussian pdf (panel (d), with $\sigma_{max}=0.3$). This definitively demonstrates that if some noise creates intermittency and  correlations, too much noise destroys them. 
\\
As further test to verify the accuracy of the q-Gaussian fit  shown in Fig.3(a), in the top panel of Fig.4 we plot (as open circles) the $q$-logarithm (defined as $\ln_q z \equiv [z^{1-q}-1]/[1-q]$, with $\ln_1 z=\ln z$)
of the corresponding pdf, normalized to its peak, as function of $x^2$, and we verify that a $q$-logarithm curve with $q=1.54$  fits very well the simulation points with a correlation coefficient equal to 0.9958. On the other hand, in the bottom panel of Fig.4,  we show that the Gaussian behavior of Fig.3(d) can be also obtained considering the same parameters of Fig.3(a),  but with the maps in the fully chaotic regime, i.e. with $\mu=2$ instead of $\mu=\mu_c$. This indicates that the edge of chaos condition is strictly necessary for the emergence of intermittency and strong correlations in presence of a small level of noise. 
Of course the cases $\mu=\mu_c$ and $\mu=2$ are two limiting ones. We also checked that, changing the order parameter in the interval $[\mu_c,2]$, other kinds of non-gaussian pdfs occurs, which are very often asymmetric or a superposition of these two extreme cases.    
Finally, in order to test   the robustness of the fat tails previously found, we show  in Fig.5 the same case reported in Fig.3(a) but now obtained with more statistics, i.e. considering $10^6$ iterations and pdfs  calculated with different number of bins. The same $q$-Gaussian 
curve of Fig.3(a) (full line)  continues to reproduce  very well the new data both in the central part and in the tails.
%
%
\begin{center}
\begin{figure}
\includegraphics[width=3.4in,angle=0]{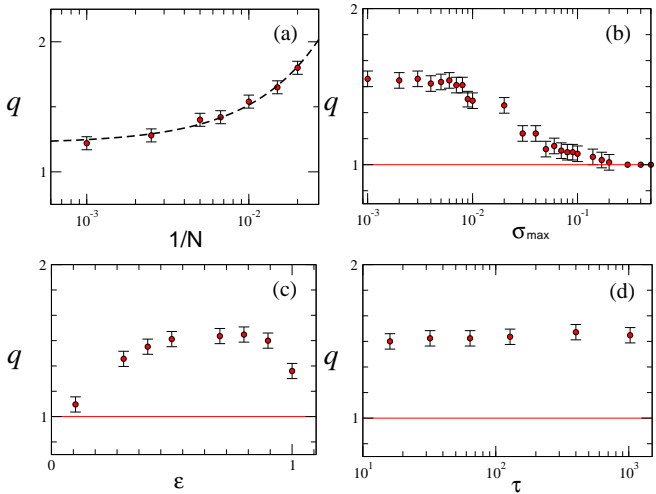}
\caption{\label{summary} 
A summary of the main results found in our study as a function, respectively, of: the size of the system $N$ (a), the level of noise $\sigma_{max}$ (b), the coupling $\epsilon$ (c) and the returns interval $\tau$ (d). The maps were always considered at the edge of chaos (where the system has a multifractal structure which is consistent with the $\tau$-independence of $q$). See text for further details. 
}
\end{figure}
\end{center}
%

If one considers the value of the entropic index $q$, emerging through $q$-Gaussian fits of the returns Pdfs, as a measure of the correlations induced by the noisy environment on our chain of coupled maps at the edge of chaos, it is worthwhile to explore how this value changes as function, not only of the noise $\sigma_{max}$, but also of the number $N$ of maps, the coupling strength $\epsilon$ and the returns time interval $\tau$.  
We show in Fig.6 a summary of the results obtained in this direction for $\mu=\mu_c=1.4011551...$ and changing the  parameters $\sigma_{max}=0.002$, $N=100$, $\epsilon=0.8$ and $\tau=32$  one  at a time and  then calculating the corresponding values of $q$ as reported. More precisely, in panel (a) we plot the entropic index as function of $1/N$ and we see that $q$ remains greater than $1$ also for very large $N$, thus implying that the noise induced correlations are not a finite-size effect. The influence of noise on the value of $q$ used to fit the pdf of the returns is reported in panel (b), where an asymptotic convergence towards $1$, for strong noise, and towards $\sim1.5$, for weak noise, is clearly visible. Quite interestingly, the plot of $q$ as function of the coupling strength, panel (c), has a maximum in correspondence of $\epsilon\sim0.8$, a value which evidently allows an optimal spreading of correlations over the maps chain in the presence of a small noise and corresponds to the fattest tails in the Pdf's, i.e.    to frequent large jumps. The fact that they occur for finite levels of $\epsilon$ is somewhat reminiscent of phenomena such as stochastic resonance \cite{Gammaitoni}. 
Finally, in panel (d) we plot $q$ versus the time interval $\tau$ used to calculate the returns: the resulting points seem to stay constant in the range of $\tau$ explored, thus confirming the robustness of this kind of correlations for low levels of noise.               
\begin{center}
\begin{figure}
\includegraphics[width=3.4in,angle=0]{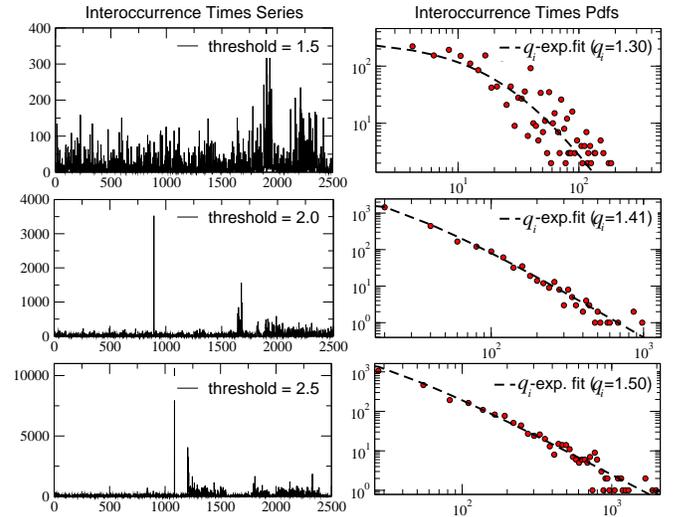}
\caption{\label{fig6} 
Left column panels: plots of the interoccurrence times $\tau_i$ of returns for increasing thresholds in the case $N=100$, $\epsilon=0.8$, $\sigma_{max}=0.002$ and $\tau=32$. Right column panels: Pdf of the time series reported on the left panels. These pdfs are nicely fitted by $q_i$-exponential  curves, whose value of $q_i$ is also reported.  See text for further details.
}
\end{figure}
\end{center}
%
%
\begin{center}
\begin{figure}
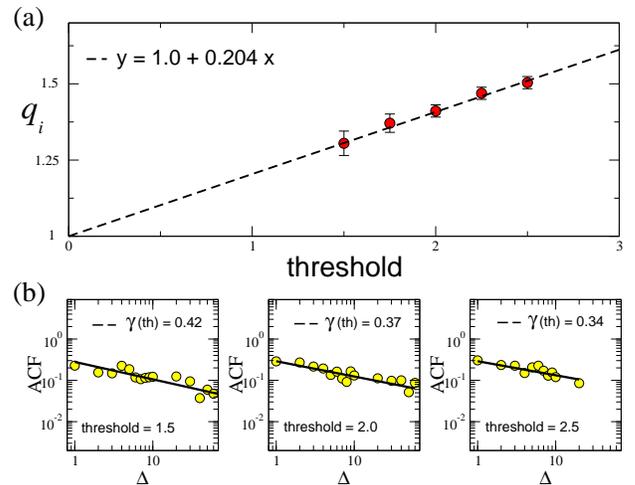

\includegraphics[width=3.2in,angle=0]{Fig8a.eps}
\includegraphics[width=3.2in,angle=0]{Fig8b.eps}
\caption{\label{fig7} 
Panel (a): The values of the index $q_i$, resulting by $q_i$-exponential fits of the interoccurrence time series Pdfs, are reported in correspondence of five values of the threshold ($1.5$, $1.75$, $2.0$, $2.25$, $2.5$). A linear fit is also plotted for comparison. Panels (b): The auto-correlation function $C_{th}(\Delta)$ for the interoccurrence time series is plotted for the three correspondent thresholds values of Fig.7 and the numerical points are fitted by power-law curves $C_{th} (\Delta) \sim \Delta^{-\gamma(th)}$ with $\gamma(th)$ equal to, respectively, $0.42$ ($th=1.5$), $0.37$ ($th=2.0$), $0.34$ ($th=2.5$).     
}
\end{figure}
\end{center}
Long-term correlations in a system typically yield power-law asymptotic behaviors in various physically relevant properties. In studies of financial markets,  it  was recently observed  \cite{Bunde} power-law decays in the so-called 'interoccurrence times' between sub sequential peaks in the fluctuating time series of returns like those shown in Fig.2(a). If we fix a given threshold, the sequence of the interoccurrence time intervals results to be well defined and it is then possible to study its pdf. We do this for our usual chain of $N=100$ maps at the edge of chaos, with $\epsilon=8$, $\tau=32$ and for a weak noise with $\sigma_{max}=0.002$. In the left panels of Fig.7, the interoccurrence time series for the normalized returns are plotted (from top to bottom) in correspondence of three increasing values of the threshold ($1.5$, $2.0$ and $2.5$), while the correspondent pdfs are reported on the right. In all the cases $q_i$-exponentials (i.e., Pdf $\propto [1-(1-q_i) \tau_i/\tau_{q_i}]^{1/1-q_i}$, where the subindex $i$ stands for {\it interoccurrence}) satisfactorily fit the data for values of $q_i$ which depend on the threshold, in complete analogy with what was observed for financial data \cite{Bunde,Bunde2}. 
\\
In Fig.8(a) we also show that $q_i$ scales linearly as function of the threshold.
This can be considered as a further footprint of the complex emergent  behavior induced on the system by the small level of noise considered.  Interestingly enough, in the limit of vanishing threshold, $q_i$ approaches unity, i.e., the behavior becomes exponential, which is precisely what was systematically observed in financial data  \cite{Bunde2}. Finally, we also calculated the auto-correlation function $C_{th}(\Delta)=\it{A'}\sum_k^{L-\Delta} {(\tau_i (k) - <\tau_i>)(\tau_i (k+\Delta) - <\tau_i>)}$ (ACF) for the interoccurrence time series reported in the left panels of Fig.7, where $L$ is the length of the time series, $th$ stands for threshold and $A'$ is  a normalization factor. As shown in Fig.8(b), for the corresponding values of the threshold considered, we found a power-law decay  $C_{th} (\Delta) \sim \Delta^{-\gamma(th)}$ with values for the exponent $\gamma(th)$ decreasing with the increase of the  threshold and included in the interval $[0.34,0.42]$, in agreement with analogous results found in financial data \cite{Bunde,Bunde2}. This shows also the presence of memory effects induced by noise, in addition to the correlations already pointed out by the deviations from Gaussian behavior quantified by the entropic index $q$.  

In conclusion, we have studied the effect of a small additive noise on a synchronized  linear chain of N   locally-coupled logistic maps at the edge of chaos. We found   strong intermittent fluctuations  in the returns, whose pdfs  are well fitted with $q$-Gaussians. The corresponding interoccurrence times for fixed threshold exhibit strong analogies with financial data. This behavior  could bring interesting  insights on the several common features of real  systems of different nature which often operate at the edge of chaos and  in weakly noisy environments. 
The study of further details of this phenomenon in various complex systems, including earthquakes, is in progress and will be reported elsewhere.

\bigskip 
Two of us, A.P. and A.R., would like to thank CBPF for the warm hospitality and the financial support received in Rio de Janeiro during the preparation of this work. We also thank Alessio Luca, Cigdem Yalcin, Salvo Rizzo and Bruno Zerbo for useful discussions. Partial financial support from CNPq and Faperj (Brazilian agencies) is also acknowledged.


\vfill
\end{document}